\documentclass[11pt]{article}
  
\usepackage{amsmath,amssymb,setspace}
\usepackage{slashed}
\usepackage{graphicx}
\usepackage{url}
\usepackage{hyperref}

\usepackage[square,numbers,comma,compress]{natbib}
\usepackage{bm,epstopdf}

\def\beq{\begin{eqnarray}}
\def\eeq{\end{eqnarray}}
\def\bea{\begin{eqnarray*}}
\def\eea{\end{eqnarray*}}
\def\lsim{\mathrel{\rlap{\lower4pt\hbox{\hskip1pt$\sim$}}
    \raise1pt\hbox{$<$}}}                
\def\gsim{\mathrel{\rlap{\lower4pt\hbox{\hskip1pt$\sim$}}
    \raise1pt\hbox{$>$}}}                

\def\mpl{M_{pl}}
\def\mgr{m_{3/2}}

\def\singleandhalfspaced{\baselineskip=\normalbaselineskip\multiply
    \baselineskip by 150\divide\baselineskip by 100}

\newcommand{\newc}{\newcommand}
\newc{\Kahler}{K\"ahler }

\begin{document}
\begin{titlepage}
\begin{flushright}
{\large SCIPP 14/10\\
}
\end{flushright}

\vskip 1.2cm

\begin{center}

{\LARGE\bf Monodromy Inflation in SUSY QCD}\\

\vskip 1.4cm

{\large  Michael Dine, Patrick Draper and Angelo Monteux }
\\
\vskip 0.4cm
{\it Santa Cruz Institute for Particle Physics and
\\ Department of Physics,
     Santa Cruz CA 95064  } \\

\vskip 4pt

\vskip 1.5cm

\vskip 1.4cm

\begin{abstract}
The discovery of a large tensor-to-scalar ratio by the BICEP2 experiment points to large field excursions during inflation.  One framework that predicts large $r$
is 
monodromy inflation. While discussed mainly in the context of string theory,
the phenomenon can be illustrated and studied in the well-understood framework of SUSY QCD with a large number of colors.  We discuss the requirements for viable inflation as well as various difficulties for model building, including tunneling, tuning, and the species problem.

\end{abstract}

\end{center}

\vskip 1.0 cm

\end{titlepage}
\setcounter{footnote}{0} \setcounter{page}{2}
\setcounter{section}{0} \setcounter{subsection}{0}
\setcounter{subsubsection}{0}

\singleandhalfspaced


\section{Introduction}
The BICEP2 discovery~\citep{Ade:2014xna,Ade:2014gua} of B-modes in the perturbations of the cosmic microwave background constitutes strong evidence for primordial gravitational waves emitted during inflation. The large value of the tensor-to-scalar ratio, $r=0.2$, provides an interesting challenge for inflationary model-building. In single-field slow-roll inflation models, a well-known argument due to Lyth~\citep{Lyth:1996im} connects a large amplitude for the tensor perturbations with transplanckian excursions in field space during inflation. Such models are challenging to place under theoretical control. 

One class of models that already pointed toward inflaton excursions beyond $\mpl$, and thus to substantial tensor
perturbations, is natural inflation~\citep{Freese:1990rb} (another is  chaotic inflation~\citep{Linde:1983gd}).   Natural inflation (NI) models rely on an approximate, spontaneously broken global symmetry to provide a quasi-flat direction, which we refer to as an axion.
Successful inflation, in the simplest version, requires an axion decay constant
substantially larger than $M_{pl}$.  Such large decay constants are hard to understand in effective field theory, particularly if quantum gravity breaks the global symmetry explicitly.  In string theory (more generally higher-dimensional theories) compactification on small volumes can appear to produce such fields~\citep{ArkaniHamed:2003wu}.  However, due to various dualities, it turns out that the effective decay constants are not parametrically large~\citep{Banks:2003sx}.   An interesting approach to evading this difficulty is monodromy inflation.  Discussed principally in string theory~\citep{Silverstein:2008sg,McAllister:2008hb}, the basic idea is to consider axions with subplanckian decay constants, $f_a < M_{pl}$, and dynamics that permit angular excursions much larger than $2\pi$.\footnote{A different approach, involving multiple axionic directions~\citep{Dimopoulos:2005ac}, will not be explored here.}
Following the BICEP2 announcement, there have been a number of studies updating and extending earlier work on natural inflation~\citep{Freese:2014nla} and monodromy inflation~\citep{Kaloper:2014zba,Marchesano:2014mla,Ibanez:2014kia,Kobayashi:2014ooa,Kallosh:2014vja}.

To fully understand monodromy inflation in string models is challenging.  The details, including issues such as moduli stabilization and tunneling from configurations on one branch to another, are inevitably quite complex.  
Monodromy inflation has also been realized in some field theory models~\citep{Kaloper:2008fb,Kaloper:2011jz,Berg:2009tg,Dubovsky:2011tu,Harigaya:2014eta,Kim:2004rp,Kappl:2014lra}. In this note we discuss a simple setting for monodromy in a familiar four-dimensional field theory, supersymmetric QCD (SQCD),  
where the number of colors is tied to the number of $e$-foldings.\footnote{Supersymmetric strong dynamics have also been studied recently in the different context of chaotic inflation~\citep{Harigaya:2014sua}.} While not necessarily advocating that a large $N$ gauge group at very high scales describes our universe, such theories provide a
theoretically tractable class of toy inflation models exhibiting the monodromy mechanism.  

In section~\ref{sec:global}, we briefly discuss globally supersymmetric QCD with a mass term for the quarks, and show that while it possesses monodromy for an angular field, it does not easily give large field excursions in that direction unless a separate SUSY-breaking sector is added. In section~\ref{sec:sugra} we couple the system to supergravity and show that adding a constant to the superpotential generates, in a manner analogous to anomaly mediation, a natural inflation potential over $\mathcal{O}(N)$ windings of the angle. Stabilizing the radial direction at subplanckian values during inflation is achieved by adding soft SUSY-breaking terms to the potential. We discuss the parameter ranges relevant for inflation and check that the hierarchies required for the validity of the effective field theory analysis can in principle be satisfied. We will see that an important shortcoming is the presence of a set of meso-tunings\footnote{Coined in~\citep{sundrum} to describe modest tunings, here of order $10^{-2}-10^{-3}$.} in the model. Finally, we discuss tunneling processes that may bring a premature end to inflation, and find that they can provide non-trivial constraints in some parameter regimes.


\section{Global SUSY}
\label{sec:global}

We start with global SQCD with one flavor and a mass term. Although we will quickly see that this theory is not suitable for large-field inflation, the failure is instructive.

On the Higgs branch, $\bar{Q}Q=\phi^2 e^{i\theta}$, the gauge symmetry is broken to $SU(N-1)$ plus a singlet, and gaugino condensation occurs with 
\begin{align}
\langle \lambda \lambda \rangle = e^{2 \pi i k \over N-1} \Lambda_L^3.
\end{align}
$\Lambda_{L}$ is the scale of the $SU(N-1)$, and is determined by $\phi$ 
($\Lambda_{L} < \phi$).
Below $\Lambda_{L}$, only the singlet remains and is governed by the ADS superpotential (essentially $\Lambda_L^3$)~\citep{Affleck:1983mk},
\begin{align}
W_{ADS}=W_{np}+m\bar{Q}{Q}=-\frac{N-1}{32\pi^2} \Lambda_{H}^3 \left(\frac{\Lambda_{H}^{2}}{\bar{Q}Q} \right)^{\frac{1}{N-1}}e^{2 \pi i k \over N-1}+m \bar{Q}Q
\end{align}
where the scales of the high and low-energy theories are related by
\begin{align}
\Lambda_{L}=\Lambda_{H}^{\frac{3N-1}{3N-3}}\phi^{-\frac{2}{3N-3}}\;.
\end{align}
We have also indicated the existence of different branches, originating from the breaking of the approximate $Z_{2(N-1)}\rightarrow Z_2$ symmetry of the low energy theory.

The appearance of the fractional power in the superpotential gives rise to monodromy for the phase $\theta$. For fixed radius $\phi$, the potential for $\theta$ contains a term
\begin{align}
V\supset \frac{1}{16\pi^2}\Lambda_L^3 m\cos\left(\frac{N\theta}{N-1}\right)\;.
\label{eq:Vglobal}
\end{align}
The angle $\theta$ here  takes values between zero and $2\pi (N-1)$. Microscopically, $\theta$ is only valued on $[0,2\pi)$.  This distinction, as is well known, arises because of the different phases of the gaugino condensate; the transformation $\theta \rightarrow \theta + 2 \pi, ~k \rightarrow k+1$ is a symmetry of the theory.
After integrating out the gaugino condensate, the branch label remains in the low energy action, effectively producing the monodromy in the ADS potential. 

\begin{figure}[h]
\begin{center}
\hspace*{-0.75cm}
\includegraphics[width=0.5\textwidth]{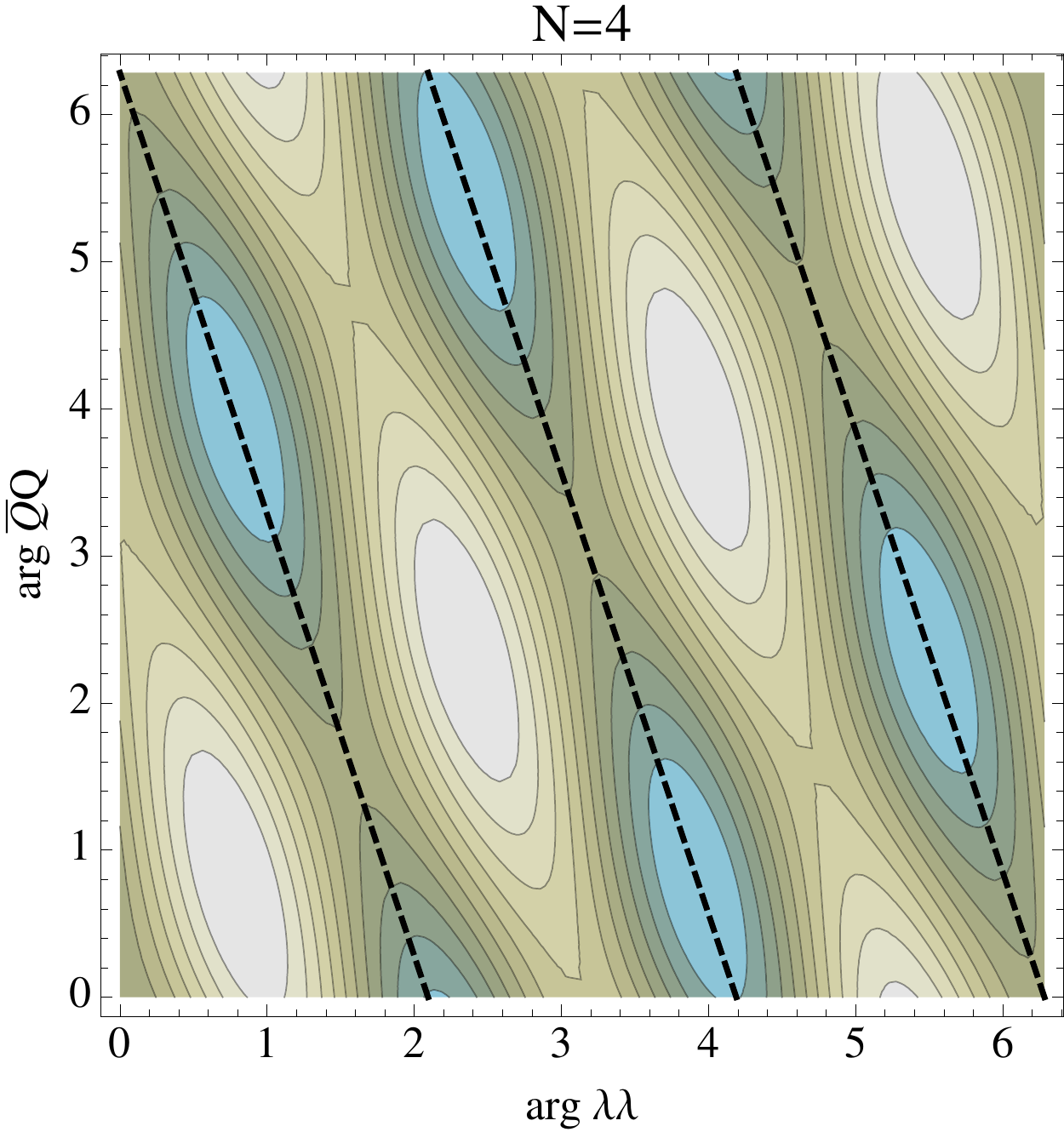} \hspace{3mm}
\vspace{0.2cm}
\caption{Schematic plot of the potential for the phases of the gaugino bilinear and the $\bar{Q}Q$ modulus in $N=4,~N_f=1$ SQCD. The dashed line is the domain of the effective theory at low energies, and it passes through multiple vacua (blue). For fixed $\bar{Q}Q$, there are $N-1=3$ branches of gaugino vacua.}
\label{fig:branches}
\end{center}
\end{figure}

This behavior is shown schematically in Fig.~\ref{fig:branches} for the case $N=4$. Microscopically, both phases are periodic with period $2\pi$, and the contours depict the qualitative behavior of the potential. The dashed line is the domain of the low-energy ADS effective theory, and it wraps three times, correspond to the three branches of gaugino vacua.  

It is clear both from Fig.~\ref{fig:branches} and Eq.~(\ref{eq:Vglobal}) that this particular exhibition of monodromy is not useful for large-field inflation. While the periodicity of $\theta$ can be $\gg2\pi$ for large $N$, the periodicity of the potential of Eq.~(\ref{eq:Vglobal}) is only $2\pi(N-1)/N$. Therefore, $\theta$ can never roll more than one cycle before encountering a potential minimum. 

To achieve large-field inflation, we want to find a $V(\theta) \sim \cos(\alpha\theta/N)$ for large $N$, where $\alpha$ is independent of $N$. This potential can generate natural inflation~\citep{Freese:1990rb}, with effective scale $f=N\phi/\alpha$. 

We can find extensions of $W_{ADS}$ in global SUSY that exhibit $\cos(\theta/N)$. They could arise from a second term in the superpotential,
\begin{align}
W=W_{ADS}+ C (\bar{Q}Q)^{-\frac{a-1}{1-N}}\;,\;\;\;a\neq1\;.
\end{align}
Such terms are suggestive of a second gauge sector also higgsed by $\bar{Q}Q$, which is rather complicated. In a simpler direction, we could instead add a sector to break supersymmetry, as in gauge mediation.  In particular, if the supersymmetry breaking is such that
$m_{\lambda} \ll \Lambda_L$, then the microscopic potential term
\begin{align}
m_{\lambda} \lambda \lambda+\rm{h.c.}
\end{align}
generates a low-energy potential
\begin{align}
V\sim m_{\lambda} \Lambda_L^3 \cos\left({\theta \over N-1}\right).
\end{align}

Such a model would be ultraviolet complete and completely calculable.  However, in the cosmological context, we must ultimately include gravity and consider the cosmological constant (now and in the inflationary epoch), and so we will give up complete calculability, coupling to gravity to linear order.  This we will do in the next section, finding in essence that the anomaly-mediated gaugino mass gives rise to the potential above. 

 To conclude this section, we note that in addition to introducing gauge-mediated supersymmetry breaking, there are other directions in which to extend the model.  One can, for example, include $N-1>N_f>1$  flavors. This extension adds light phase degrees of freedom while reducing monodromy, $\cos(\theta/(N-1))\rightarrow\cos(\theta/(N-N_f))$, and we do not pursue it further here.


\section{Supergravity}
\label{sec:sugra}
\subsection{Anomaly-mediated model}
The models discussed in the previous section illustrate that monodromy of the sort needed for inflation is a familiar phenomenon in field theory.  In order to implement a model of inflation, it is necessary to couple to (super)gravity, and this coupling introduces new possibilities for model building.  For the simplest example, physics is now sensitive to the constant in the superpotential. We will take
\begin{align}
W=W_{np}+W_0\;.
\end{align}
with $\langle W\rangle\approx W_0$, so that $\mgr\approx W_0/\mpl^2$. Then from the $-3|W|^2$ term in supergravity potential, we obtain a term 
\begin{align}
V\supset -6 |W_{np}| m_{3/2} \cos(\theta/N)\;,
\label{eq:costhetaN}
\end{align}
where we have assumed $N$ is large so that $N-1\approx N$ and $|W_{np}|$ is approximately a constant proportional to $\Lambda_H^3$. In this potential the anomaly-mediated gaugino mass\footnote{Since $W_{np}$ is proportional to the gluino condensate, identifying the $W_{np}W_0^*$ term in the low-energy potential provides an effective field theory derivation of the anomaly-mediated gaugino mass in the microscopic theory~\citep{Dine:2013nka}.} has taken the role of the gauge-mediated gaugino mass discussed in the previous section. To simplify notation, henceforth we will use $W_{np}$ to denote the modulus of its scalar component, showing phase dependence explicitly and setting the bare theta-angle to zero. We also add SUSY-breaking terms to the potential, parametrized by 
\begin{align}
V_{SB}=\lambda_0' \mgr^2\mpl^2 + \frac{1}{2}\lambda_2' \mgr^2 (\bar{Q}^\dagger\bar{Q}+Q^\dagger Q)+ \frac{1}{4}\left(\frac{\mgr}{\mpl}\right)^2 (\bar{Q}^\dagger\bar{Q}+Q^\dagger Q)^2\;.
\label{eq:vsb}
\end{align}
The primary purpose of the SUSY-breaking terms is to set a zero of the potential and to stabilize the radial direction $|\bar{Q}Q|$.  Other choices for $V_{SB}$ are possible, and even here we have not introduced as many parameters as we might.  The main point is that the potential will be minimized with $\bar Q Q $ somewhat less than $M_{pl}^2$; we choose the form of Eq.~(\ref{eq:vsb}) for simplicity.

We can easily write down theories with additional fields that spontaneously break SUSY and give rise to $V_{SB}$.  For example, one can add terms of the form $f X$ to $W$ and $X^\dagger X(1+\lambda/\mpl^2 Q^\dagger Q)$ to $K$. 
In any case, replacing $\bar{Q}^\dagger\bar{Q},Q^\dagger Q\rightarrow \phi^2$, the leading terms in the potential have the form
\begin{align}
V=\lambda_0\mgr^2\mpl^2 + \lambda_2 \mgr^2 \phi^2+ \left(\frac{\mgr}{\mpl}\right)^2 \phi^4-6W_{np} \mgr\cos\left({\theta/N}\right)+\mathcal{O}\left(\frac{W_{np}^2}{\mpl^2}\right)\;
\label{fullpotential}
\end{align}
where the unprimed coefficients differ from the primed coefficients in~(\ref{eq:vsb}) by order-1 contributions from the SUSY-preserving potential.

With this potential, for $\lambda_2<0$, the radial direction is stabilized at a large but subplanckian value for small $|\lambda_2|$,
\begin{align}
\phi\approx \sqrt{|\lambda_2|/2}\mpl\;.
\end{align}
Its mass is of order $m_\phi^2 \sim |\lambda_2| \mgr^2$, so $|\lambda_2|$ cannot be too small if $\phi$ is to be frozen during inflation.  Moreover, $\phi$ cannot be too small without requiring an enormous value for $N$. We return to these and other conditions on the parameters in~Sec.~\ref{sec:conditions}.

The potential for $\theta$, integrating out $\phi$ and tuning $\lambda_0$ so that the cosmological constant vanishes when the cosine is at
its maximum, is given by
\begin{align}
V(\theta) = 6W_{np} m_{3/2} \left[1- \cos\left({\theta / N }\right)\right]\;.
\label{vcosine}
\end{align}

\subsection{Inflation}

In the gauge theories of inflation under consideration here, $1/N$ will act as the small parameter responsible for slow-roll.   In order to obtain a sufficient number of $e$-foldings, inflation typically begins when $\theta \sim 3N \pi/4$. It ends when the slow-roll parameters become of order one, which occurs near the minimum -- zero -- of the potential, when $\theta \approx 0$.

$V(\theta)$ has exactly the form of the natural inflation potential, $V=\Lambda^4(1-\cos(a/f))$, with a product of scales playing the role of the dynamical scale $\Lambda$ and an effective symmetry breaking scale $f$,
\begin{align}
\Lambda^4\equiv6W_{np}\mgr\;,\;\;\;\;f\equiv N\phi.
\end{align}
The relevant magnitudes for these parameters and the initial conditions are well-known. We review them here in brief. For a precise study and detailed review of natural inflation in light of BICEP2, see~\citep{Freese:2014nla}, in particular Figs. 1 and 5 (note that for comparison, the reduced Planck mass used in this work must be converted to the Planck mass used in~\citep{Freese:2014nla}). 

The slow-roll parameters take the NI form,
\begin{align}
\epsilon=\frac{\mpl^2}{2N^2\phi^2}\frac{\sin^2(\theta/N)}{(1-\cos(\theta/N))^2}\;,\;\;\;\;\eta=\frac{\mpl^2}{N^2\phi^2}\frac{\cos(\theta/N)}{1-\cos(\theta/N)}\;.
\end{align}
At generic points, both $\epsilon$ and $\eta$ are of order $(\mpl/N\phi)^2$, so $f>\mpl$ is necessary in NI.  In that case $\epsilon\rightarrow 1$ shortly before $\eta\rightarrow1$. The value of $\theta$ when inflation ends is given by
\begin{align}
\cos(\theta_{\rm end}/N)=\frac{2f^2-\mpl^2}{2f^2+\mpl^2}\;,
\end{align}
while the value of $\theta$ at $N_{\rm CMB}\sim60$ e-folds before the end of inflation is determined by
\begin{align}
\cos(\theta_{\rm CMB}/N)=\frac{4 f^2}{2f^2+\mpl^2}e^{-N_{\rm CMB} \mpl^2/f^2}-1\;.
\end{align}
As usual the overall scale of the potential and the effective symmetry breaking scale $f$ are constrained by the amplitude of perturbations and the scalar spectral index. Using the relation above for $\theta_{\rm CMB}$, the spectral index is approximated by
\begin{align}
n_s&=1+2\eta_{\rm CMB}-6\epsilon_{\rm CMB}\nonumber\\
&=1-\frac{\mpl^2}{f^2}-\frac{4}{e^{N_{\rm CMB} \mpl^2/f^2}(1+2f^2/\mpl^2)-2f^2/\mpl^2}\;
\end{align}
from which we find $f\gtrsim \mathcal{O}(10)\mpl$ for $n_s=0.96$ ($f\approx10$ when $N_{\rm CMB}=50$). Similarly, the amplitude of density perturbations constrains
\begin{align}
\Delta_s^2&=\frac{V^3}{12\pi^2\mpl^6V'^2}\nonumber\\
&=\frac{f^2 \Lambda^4 (1-\cos(\theta_{\rm CMB}/N))^3}{12\pi^2\mpl^6\sin^2(\theta_{\rm CMB}/N)}\nonumber\\
&\sim10^{-9}\;,
\end{align}
which, for $N_{CMB}\sim60$ and $f\sim10M_{pl}$, fixes
\begin{align}
\Lambda= 1.8\times10^{16} {\rm~GeV}\;.
\end{align}
 This corresponds to $r\approx0.1$, and is consistent with the BICEP2 measurement of large $r$ within uncertainties.\footnote{At present there appears to be some tension between Planck~\citep{Ade:2013uln} and BICEP2~\citep{Ade:2014xna}. The value of $r$ cited by BICEP2  after subtracting estimated foreground dust contributions is $r=0.16$. A joint likelihood fit between Planck and BICEP2 suggests $r\approx 0.15\pm.05$~\citep{Freese:2014nla}.
}
It is difficult to achieve $r=0.2$ in natural inflation as it requires both large $f$ and a small number of $e$-foldings, $N_{CMB}\sim 40$. 

\subsection{Self-Consistency}
\label{sec:conditions}
Establishing various hierarchies is critical for the internal self-consistency of the analysis. Consistency of the supergravity, $SU(N-1)$, and moduli effective theories requires approximate hierarchies of the form
\begin{align}
\phi/\mpl\ll 1\;,\;\;\;\;\Lambda_L/\phi \ll 1\;,\;\;\;\; \mgr/\Lambda_L \ll 1\;.
\end{align}
Furthermore, we ignored $W_{np}^2/\mpl^4$ contributions to the $\phi$ mass term in the potential. Therefore, we require
\begin{align}
\frac{W_{np}^2}{\mgr^2\mpl^4}\ll 1\;.
\label{eq:constr1}
\end{align}
To stabilize the radial direction during inflation, we need
\begin{align}
\frac{H^2}{m_\phi^2}=\frac{W_{np}}{|\lambda_2|m_{3/2}\mpl^2}\ll1\;,
\end{align}
which implies~(\ref{eq:constr1}) when $\lambda_2$ is small. All conditions can be satisfied. For example, if the parameters numerically satisfy
\begin{align}
|\lambda_2|&\sim 1/N\;,\nonumber\\
W_{np}^{1/3}/\mpl&\sim 1/N\;,\nonumber\\
\mgr/W_{np}^{1/3}&\sim 1/\sqrt{N}\;,
\end{align}
 the theory is self-consistent for large $N$. In this case, $\phi/\mpl\sim 1/\sqrt{N}$ and $N\phi/\mpl\sim \sqrt{N}$, and the energy scale of inflation is of order $N^{-9/2}\mpl^4$. Therefore, $N\sim100$, and all hierarchies are satisfied by an order of magnitude or more.\footnote{Note that $W_{np}\sim \Lambda_L^3$ for $N\sim 100$. }

Needless to say, a gauge group such as $SU(100)$ at a very high energy scale does not seem a particularly plausible model
of nature.  Despite these issues, the virtue of the setup is that it provides a simple field-theoretic realization of monodromy inflation.
 
Of course, setting aside phenomenology, other scalings are possible which appear to permit arbitrarily large $N$. For example, one could take $\phi/\mpl\sim10^{-2}$, $\Lambda_{L}/\mpl\sim N^{-3}$, $\Lambda_{L}/\mgr\sim N^{-2}$. However, as we will discuss in the next section, there are fine-tunings in the model, and even without imposing phenomenological constraints, some these tunings grow parametrically with $N$.

\subsection{Tunings and Higher Order Corrections}

In its simplest version, natural inflation requires large $f_a$ and a softly broken, presumably accidental, shift symmetry.
In conventional field theory, these requirements are challenging.  In string theory, the accidental symmetries can result from discrete shift symmetries, but, as noted previously, large $f_a$ does not seem to arise easily.  Monodromy inflation avoids the difficulty.  Still, in an effective field theory framework, one might expect that there are effects that can spoil the story. Potential problems include the renormalization of Newton's constant, the destabilization of the $\phi$ vacuum by higher dimension operators, and the violation of slow-roll conditions, satisfied by low dimension terms in the action, by  higher dimension operators.

As emphasized in~\citep{ArkaniHamed:2005yv,Distler:2005hi,Dimopoulos:2005ac}, $\mpl^2$ receives quadratically divergent renormalization, which can be large in the presence of a large number of degrees of freedom and if the cutoff is of order $\mpl$.  This casts doubt on the validity of effective field theory.  One form of the problem is the potential need to fine-tune the bare Planck mass at the UV cutoff scale, $\Lambda_{UV}$.    In terms of the bare mass ${\mpl^2}_0$ and the number of fields $\tilde{N}$, the effective Planck mass is
\begin{align}
\mpl^2\approx {\mpl^2}_0 \pm \frac{\tilde{N}\Lambda_{UV}^2}{16\pi^2}\;,\;\;\;\;\;\tilde{N}\sim N^2\;.
\end{align}
If $\Lambda_{UV}\sim \mpl$, then ${\mpl^2}_0$ must be tuned to a part in $(N/4\pi)^2$. This sort of percent-level fine-tuning is an unappealing feature of the model. For fixed $\Lambda_{UV}$, the tuning is parametrically worsened with $N$. We could lower $\Lambda_{UV}$ to decrease the tuning, but we would simultaneously have to reduce $\phi$, which would in turn require that we further increase $N$ (and ultimately become inconsistent with the phenomenology of inflation).  
For the rest of this section we assume $\Lambda_{UV}=\mpl$.

Returning to the potential~(\ref{vcosine}), dangerous symmetry-breaking operators are suppressed by $(\phi/\mpl)^p<1$. However, we have seen that $\phi/\mpl$ cannot be arbitrarily small. Therefore, we may expect that we require some additional number of fine-tunings in order to achieve both the stabilization of the $\phi$ potential and the suppression of operators that would spoil the flatness of the $\theta$ potential.
 
In the $\phi$ potential, Eq.~(\ref{fullpotential}), the tuning of the parameter $\lambda_0$ is the usual tuning of the cosmological constant; in natural inflation,
this tuning also allows inflation to end. The parameter $\lambda_2$ must be small to allow $\phi<\mpl$. In the example above, $|\lambda_2|\sim 1/N$ implies a tuning of the $\phi$ mass parameter of a part in $N$. On the other hand, terms of order $(Q^\dagger Q)^2/\mpl^2$ in the \Kahler potential provide innocuous corrections once we require the hierarchies listed in the previous section.

In the absence of symmetries or some other microscopic considerations, we should include a variety of terms in the effective theory, among them
\begin{align}
\mathcal{O}_K=Q^\dagger Q \bar{Q}Q/\mpl^2
\end{align}
in the \Kahler potential, and
\begin{align}
\mathcal{O}^p_W=(\bar{Q}Q)^p \mpl^{3-2p}\;,\;\;\;\;p=2,3,4...
\end{align}
in the superpotential.
$\mathcal{O}_K$ contributes to the inflaton potential, giving rise to terms of order $V\phi^2/N\mpl^2~\cos(\theta)$ and $V\phi^4/\mpl^4~\cos(\theta)$.   Such terms do not exhibit monodromy, but may be sufficiently suppressed by powers of $N$ and $\phi$.
 The $\mathcal{O}^p_W$, for $p\sim2,3$, are potentially more problematic. The $-3|W|^2$ part of the potential introduces 
\begin{align}
V\supset\alpha_p \frac{m_{3/2}\phi^{2p}}{\mpl^{2p-3}}\cos(p\theta)
\end{align}
 terms in the large $N$ limit, and these must be small compared to the $\mgr W_{np}\cos(\theta/N)$ term. For $p\gtrsim3$, the $\alpha_p$ can be $\mathcal{O}(1)$, but for $p\sim2,3$ we require $\alpha_p$ to be small,
 \begin{align}
 \alpha_p\ll \frac{W_{np}}{\mpl^3}\cdot\left(\frac{\mpl}{\phi}\right)^{2p}\;.
 \end{align}
For the example scalings in the previous section, we require $\alpha_2\lesssim10^{-3}$ and $\alpha_3\lesssim10^{-1}$ in order for the $\cos(p\theta)$ corrections to be negligible.

Both the problematic symmetry-breaking terms in the \Kahler potential and the superpotential can in principle be suppressed by discrete symmetries (we do not expect continuous global symmetries in theories of quantum gravity).  These symmetries do not need to particularly elaborate to provide adequate suppression.

\subsection{Tunneling}
As the field $\theta$ rolls, except for very small $\theta$, in addition to rolling, the system can tunnel to a different branch with lower energy.  This is analogous to tunneling processes which have been discussed for stringy monodromy inflation. In SQCD, the tunneling amplitudes are easily estimated.  Whether they are sufficiently small to permit inflation depends on the parameters.

We are used to the idea that tunneling amplitudes are small, because treated semiclassically, the bounce action is often large.  It has been noted, however, that there can be {\it substantial} suppression of the exponent in tunneling between gaugino vacua in SQCD~\citep{Dvali:1996xe,Kovner:1997ca,Dvali:1999pk}.  We have to check if the bounce action remains large in the regime of parameters relevant for inflation.  
The bounce action can be determined in terms of the parameters $W_{np}$, $\mgr$, and $N$, and is given by~\citep{Coleman:1977py}
\beq
S_b =  {27 \pi^2T^4 \over 2\epsilon^3},
\eeq
where $T$ is the bubble wall tension and $\epsilon$ is the energy splitting between the gaugino vacua. The tension is determined by twice the modulus of the difference in the superpotential between the two vacua~\citep{Kovner:1997ca}. Recalling that 
\beq
W = W_{np} e^{i \theta/N} + W_0
\eeq
and that $\theta\rightarrow\theta+2\pi$ shifts the branch of the vacuum by one, 
the tension is
\beq
T = 2W_{np} \cdot{2 \pi \over N}.
\eeq
The energy splitting between adjacent vacua is controlled by the $\cos(\theta/N)$ term in the potential, Eq.~(\ref{eq:costhetaN}). Therefore we find
\beq
{\Delta E =6  W_{np} \mgr \sin({\theta / N})} \cdot {2 \pi \over N}\;.
\eeq
Putting the terms together,
\beq
S_b \geq {2\pi^3 \over N} { W_{np} \over   \mgr^3}\;.
\eeq
Taking the BICEP2 result for the energy scale of inflation~\citep{Ade:2014xna}, 
\beq
\mgr W_{np} \approx 10^{-9} \mpl^4\;.
\eeq
Requiring $m_\phi^2\geq H^2\sim V/M_{pl}^2$ during inflation gives a lower bound on $\mgr$.   Introducing a parameter
$\varepsilon <1$, we can write
\beq
m_{3/2}^2 = {1 \over \varepsilon}  \left ({W_{np} m_{3/2} \over M_{pl}^2} \right ) \simeq {1 \over \varepsilon} 10^{-9}M_{pl}^2.
\eeq
Then
 \begin{align}
S_b \geq {2\pi^3 \over N} {10^{9}} \varepsilon^{2}\;.
\end{align}
Finally, from the hierarchy $\mgr< W_{np}^{1/3}$ and the relation $m_\phi^2=\lambda_2 \mgr^2$, we find bounds on $\varepsilon$,
\beq
10^{-9/2}<\varepsilon<\lambda_2\;.
\eeq
So we see that tunneling may or may not be an issue; there is the possibility of significant enhancement to the rate for small $\varepsilon$, while for larger values tunneling may be sufficiently suppressed. For the sample scalings given in the previous section, $\varepsilon$ is of size $N^{-3/2}\sim10^{-3}$, giving $S_b\sim16\pi^3$. The suppression in this case is enough to permit inflation.

\section{Conclusions}
\label{sec:concl}
Monodromy is an implementation of natural inflation that avoids superplanckian axion decay constants, problematic both in field theory and string theory.  In this note we have described a simple setting for monodromy inflation in a toy field theory. In single-flavor supersymmetric QCD with a large number of colors, the supergravity potential at low energies contains an angular degree of freedom valued on $[0,2\pi N)$. Soft SUSY-breaking can stabilize the radial degree of freedom at a value below $\mpl$, while the effective field excursion of the angular variable can be transplanckian, as needed to produce a large tensor-to-scalar ratio. While we stress that this model is unlikely to reflect nature, it provides a simple class of field theory models exhibiting viable monodromy inflation, and perhaps can be used to study properties of the mechanism.  In addition to realizing the slow-roll requirements, it provides a setting in which issues including the constraints on quantum and gravitational corrections and tunneling processes are readily analyzed.  Percent-level tunings are required to control the Newton constant and the axion decay constant, while global symmetry-breaking corrections to the potential can be controlled by modest discrete symmetries, and tunneling provides interesting but not prohibitive constraints on parameters.

\vspace{1cm}

\noindent

{\bf Acknowledgements:}  This work was supported by the U.S. Department of Energy grant number DE-FG02-04ER41286.  We thank Tom Banks for discussions, and in particular for reminding us of the potential importance of $N^2$ enhanced corrections.  We also acknowledge discussions with Laurel Stephenson-Haskins.

\bibliographystyle{apsrev4-1}
\bibliography{monodromy}{}

\end{document}